\documentclass[twocolumn, superscriptaddress, amsmath,amssymb, prc aps]{revtex4-2}

\usepackage{romanbar}
\usepackage{graphicx}
\usepackage{dcolumn}
\usepackage{bm}
\usepackage{array}
\usepackage{graphicx}  
\usepackage{siunitx}  
\usepackage{epsfig} 
\usepackage{subfigure}
\usepackage{lineno}  
\usepackage{amsmath,amssymb}
\usepackage{comment}
\usepackage{multirow}
\usepackage{dcolumn}
\usepackage{bm}
\usepackage{setspace}
\includecomment{printrobustnotes}
\includecomment{printallnotes}
\usepackage{xspace}
\usepackage{color}
\usepackage{hyperref}
\hypersetup{colorlinks=true,linkcolor=blue, linktocpage}
\usepackage{siunitx}
\usepackage{array}
\usepackage{tabularx}
\usepackage{booktabs}
\usepackage{url}
\usepackage{textcomp}
\usepackage{tikz-feynman}
\tikzfeynmanset{compat=1.1.0}

\begin{document}

\title{Constraints on sub-GeV dark matter scattering on electrons with COSINE-100}
\author{N.~Carlin}
\affiliation{Physics Institute, University of S\~{a}o Paulo, 05508-090, S\~{a}o Paulo, Brazil}
\author{J.~Y.~Cho}
\affiliation{Department of Physics, Kyungpook National University, Daegu 41566, Republic of Korea}
\author{S.~J.~Cho}
\affiliation{Center for Underground Physics, Institute for Basic Science (IBS), Daejeon 34126, Republic of Korea}
\author{S.~Choi}
\affiliation{Department of Physics and Astronomy, Seoul National University, Seoul 08826, Republic of Korea} 
\author{A.~C.~Ezeribe}
\affiliation{Department of Physics and Astronomy, University of Sheffield, Sheffield S3 7RH, United Kingdom}
\author{L.~E.~Fran{\c c}a}
\email{luis.eduardo.franca@usp.br}
\affiliation{Physics Institute, University of S\~{a}o Paulo, 05508-090, S\~{a}o Paulo, Brazil}
\author{O.~Gileva}
\affiliation{Center for Underground Physics, Institute for Basic Science (IBS), Daejeon 34126, Republic of Korea}
\author{C.~Ha}
\affiliation{Department of Physics, Chung-Ang University, Seoul 06973, Republic of Korea}
\author{I.~S.~Hahn}
\affiliation{Center for Exotic Nuclear Studies, Institute for Basic Science (IBS), Daejeon 34126, Republic of Korea}
\affiliation{Department of Science Education, Ewha Womans University, Seoul 03760, Republic of Korea} 
\affiliation{IBS School, University of Science and Technology (UST), Daejeon 34113, Republic of Korea}
\author{S.~J.~Hollick}
\affiliation{Department of Physics and Wright Laboratory, Yale University, New Haven, CT 06520, USA}
\author{E.~J.~Jeon}
\affiliation{Center for Underground Physics, Institute for Basic Science (IBS), Daejeon 34126, Republic of Korea}
\affiliation{IBS School, University of Science and Technology (UST), Daejeon 34113, Republic of Korea}
\author{H.~W.~Joo}
\affiliation{Department of Physics and Astronomy, Seoul National University, Seoul 08826, Republic of Korea} 
\author{W.~G.~Kang}
\affiliation{Center for Underground Physics, Institute for Basic Science (IBS), Daejeon 34126, Republic of Korea}
\author{M.~Kauer}
\affiliation{Department of Physics and Wisconsin IceCube Particle Astrophysics Center, University of Wisconsin-Madison, Madison, WI 53706, USA}
\author{B.~H.~Kim}
\affiliation{Center for Underground Physics, Institute for Basic Science (IBS), Daejeon 34126, Republic of Korea}
\author{D.~Y.~Kim}
\affiliation{Center for Underground Physics, Institute for Basic Science (IBS), Daejeon 34126, Republic of Korea}
\author{H.~J.~Kim}
\affiliation{Department of Physics, Kyungpook National University, Daegu 41566, Republic of Korea}
\author{J.~Kim}
\affiliation{Department of Physics, Chung-Ang University, Seoul 06973, Republic of Korea}
\author{K.~W.~Kim}
\affiliation{Center for Underground Physics, Institute for Basic Science (IBS), Daejeon 34126, Republic of Korea}
\author{S.~H.~Kim}
\affiliation{Center for Underground Physics, Institute for Basic Science (IBS), Daejeon 34126, Republic of Korea}
\author{S.~K.~Kim}
\affiliation{Department of Physics and Astronomy, Seoul National University, Seoul 08826, Republic of Korea}
\author{W.~K.~Kim}
\affiliation{IBS School, University of Science and Technology (UST), Daejeon 34113, Republic of Korea}
\affiliation{Center for Underground Physics, Institute for Basic Science (IBS), Daejeon 34126, Republic of Korea}
\author{Y.~D.~Kim}
\affiliation{Center for Underground Physics, Institute for Basic Science (IBS), Daejeon 34126, Republic of Korea}
\affiliation{IBS School, University of Science and Technology (UST), Daejeon 34113, Republic of Korea}
\author{Y.~H.~Kim}
\affiliation{Center for Underground Physics, Institute for Basic Science (IBS), Daejeon 34126, Republic of Korea}
\affiliation{IBS School, University of Science and Technology (UST), Daejeon 34113, Republic of Korea}
\author{B.~R.~Ko}
\affiliation{Department of Accelerator Science, Korea University, Sejong 30019, Republic of Korea}
\author{Y.~J.~Ko}
\affiliation{Department of Physics, Jeju National University, Jeju 63243, Republic of Korea}
\author{D.~H.~Lee}
\affiliation{Department of Physics, Kyungpook National University, Daegu 41566, Republic of Korea}
\author{E.~K.~Lee}
\affiliation{Center for Underground Physics, Institute for Basic Science (IBS), Daejeon 34126, Republic of Korea}
\author{H.~Lee}
\affiliation{IBS School, University of Science and Technology (UST), Daejeon 34113, Republic of Korea}
\affiliation{Center for Underground Physics, Institute for Basic Science (IBS), Daejeon 34126, Republic of Korea}
\author{H.~S.~Lee}
\affiliation{Center for Underground Physics, Institute for Basic Science (IBS), Daejeon 34126, Republic of Korea}
\affiliation{IBS School, University of Science and Technology (UST), Daejeon 34113, Republic of Korea}
\author{H.~Y.~Lee}
\affiliation{Center for Exotic Nuclear Studies, Institute for Basic Science (IBS), Daejeon 34126, Republic of Korea}
\author{I.~S.~Lee}
\affiliation{Center for Underground Physics, Institute for Basic Science (IBS), Daejeon 34126, Republic of Korea}
\author{J.~Lee}
\affiliation{Center for Underground Physics, Institute for Basic Science (IBS), Daejeon 34126, Republic of Korea}
\author{J.~Y.~Lee}
\affiliation{Department of Physics, Kyungpook National University, Daegu 41566, Republic of Korea}
\author{M.~H.~Lee}
\affiliation{Center for Underground Physics, Institute for Basic Science (IBS), Daejeon 34126, Republic of Korea}
\affiliation{IBS School, University of Science and Technology (UST), Daejeon 34113, Republic of Korea}
\author{S.~H.~Lee}
\affiliation{IBS School, University of Science and Technology (UST), Daejeon 34113, Republic of Korea}
\affiliation{Center for Underground Physics, Institute for Basic Science (IBS), Daejeon 34126, Republic of Korea}
\author{S.~M.~Lee}
\affiliation{Department of Physics and Astronomy, Seoul National University, Seoul 08826, Republic of Korea} 
\author{Y.~J.~Lee}
\affiliation{Department of Physics, Chung-Ang University, Seoul 06973, Republic of Korea}
\author{D.~S.~Leonard}
\affiliation{Center for Underground Physics, Institute for Basic Science (IBS), Daejeon 34126, Republic of Korea}
\author{N.~T.~Luan}
\affiliation{Department of Physics, Kyungpook National University, Daegu 41566, Republic of Korea}
\author{V.~H.~A.~Machado}
\affiliation{Physics Institute, University of S\~{a}o Paulo, 05508-090, S\~{a}o Paulo, Brazil}
\author{B.~B.~Manzato}
\affiliation{Physics Institute, University of S\~{a}o Paulo, 05508-090, S\~{a}o Paulo, Brazil}
\author{R.~H.~Maruyama}
\affiliation{Department of Physics and Wright Laboratory, Yale University, New Haven, CT 06520, USA}
\author{S.~L.~Olsen}
\affiliation{Center for Underground Physics, Institute for Basic Science (IBS), Daejeon 34126, Republic of Korea}
\author{H.~K.~Park}
\affiliation{Department of Accelerator Science, Korea University, Sejong 30019, Republic of Korea}
\author{H.~S.~Park}
\affiliation{Korea Research Institute of Standards and Science, Daejeon 34113, Republic of Korea}
\author{J.~C.~Park}
\affiliation{Department of Physics and Institute for Sciences of the Universe, Chungnam National University, Daejeon 34134, Republic of Korea}
\author{J.~S.~Park}
\affiliation{Department of Physics, Kyungpook National University, Daegu 41566, Republic of Korea}
\author{K.~S.~Park}
\affiliation{Center for Underground Physics, Institute for Basic Science (IBS), Daejeon 34126, Republic of Korea}
\author{K.~Park}
\affiliation{Center for Underground Physics, Institute for Basic Science (IBS), Daejeon 34126, Republic of Korea}
\author{S.~D.~Park}
\affiliation{Department of Physics, Kyungpook National University, Daegu 41566, Republic of Korea}
\author{R.~L.~C.~Pitta}
\affiliation{Physics Institute, University of S\~{a}o Paulo, 05508-090, S\~{a}o Paulo, Brazil}
\author{H.~Prihtiadi}
\affiliation{Department of Physics, Universitas Negeri Malang, Malang 65145, Indonesia}
\author{S.~J.~Ra}
\affiliation{Center for Underground Physics, Institute for Basic Science (IBS), Daejeon 34126, Republic of Korea}
\author{C.~Rott}
\affiliation{Department of Physics and Astronomy, University of Utah, Salt Lake City, UT 84112, USA}
\author{K.~A.~Shin}
\affiliation{Center for Underground Physics, Institute for Basic Science (IBS), Daejeon 34126, Republic of Korea}
\author{D.~F.~F.~S. Cavalcante}
\affiliation{Physics Institute, University of S\~{a}o Paulo, 05508-090, S\~{a}o Paulo, Brazil}
\author{M.~K.~Son}
\affiliation{Department of Physics and Institute for Sciences of the Universe, Chungnam National University, Daejeon 34134, Republic of Korea}
\author{N.~J.~C.~Spooner}
\affiliation{Department of Physics and Astronomy, University of Sheffield, Sheffield S3 7RH, United Kingdom}
\author{L.~T.~Truc}
\affiliation{Department of Physics, Kyungpook National University, Daegu 41566, Republic of Korea}
\author{L.~Yang}
\affiliation{Department of Physics, University of California San Diego, La Jolla, CA 92093, USA}
\author{G.~H.~Yu}
\affiliation{Center for Underground Physics, Institute for Basic Science (IBS), Daejeon 34126, Republic of Korea}
\collaboration{COSINE-100 Collaboration}

\date{\today}

\begin{abstract}
We present results of the search for sub-GeV dark matter interaction with electrons in the NaI(Tl) crystals of the COSINE-100 experiment. The two benchmark scenarios of a heavy and a light vector boson as mediator of the interaction were studied. We found no excess events over the expected background in a data-set of 2.82 years, with a total exposure of 172.9 kg-year. The derived 90\% confidence level upper limits exclude a DM-electron scattering cross section above 6.4 $\times$ 10$^{-33}$ cm$^2$ for a DM mass of 0.25 GeV, assuming a light mediator; and above 3.4 $\times$ 10$^{-37}$ cm$^2$ for a 0.4 GeV DM, assuming a heavy mediator, and represent the most stringent constraints for a NaI(Tl) target to date. We also briefly discuss a planned analysis using an annual modulation method below the current 0.7 keV threshold of COSINE-100, down to few photoelectrons yield.
\end{abstract}

\maketitle

\section{Introduction}\label{sec:introduction}
The existence of cold dark matter (CDM) is supported by numerous astrophysical and cosmological observations that demonstrate its gravitational interactions \cite{dmevidences1,dmevidences2,dmreview}. All evidence is compatible with 85\% of the matter in the universe being composed of DM. Nevertheless, the precise nature of DM particles remains unknown, prompting extensive research in the field of particle physics \cite{cdm,hdm,wdm,theories}. One of the main DM candidates is the Weakly Interacting Massive Particle (WIMP). Due to the WIMP miracle \cite{paradigm} and since different theories that extend the standard model propose particles with the WIMP properties, WIMPs are rendered as a highly promising candidate \cite{wimp}. 

To date, the most studied scenario considers a weak interaction between WIMPs and nucleons. Nonetheless, because there is no compelling evidence of WIMP detection in the results from multiple direct detection experiments utilizing different detector materials, as well as the increasingly stringent constraints on the WIMP-nucleon cross section and WIMP mass \cite{cresst,darksidewimpnucleon,LZ,xenon1tmigdal}, studies of alternative models are encouraged \cite{boosted,milli,dp,axion}. One such alternative involves interactions between WIMP-like cold DM and electrons \cite{wimpelectronandnucleon}, which could be detected in modern low threshold detectors. This is particularly relevant in the context of sub-GeV particles, where increased sensitivity to this interaction is anticipated. Given that the largest unexplored region pertains to sub-GeV WIMP-like DM, the search for the scattering of this light DM and electrons is a viable and appropriate scenario.

Several collaborations have reported their searches for the light DM-electron interaction, constraining its cross section \cite{xenon1twimpelectron,xenon10wimpelectron,cdex10,edelweisswimpelectron}. As a consequence of the low threshold, silicon based experiments, such as DAMIC-M \cite{damicmwimpelectron} and SENSEI \cite{senseiwimpelectron}, have set the most stringent limits for sub-GeV DM within the mass region of 1 MeV to approximately 50 MeV. In contrast, as a result of the high detector exposure and low background, liquid noble gas experiments, such as PANDAX-4T \cite{pandaxwimpelectron}, DarkSide-50 \cite{darksidewimpelectron} and XENONnT \cite{xenonntwimpelectron}, have provided the strongest constraints for light DM with masses exceeding 50 MeV. Despite that, up to this moment, no experiment utilizing NaI(Tl) detectors has published a dedicated search for this cold sub-GeV DM-electron interaction model. The NEON experiment has recently published a search for light DM scattering on electrons in NaI(Tl) detectors \cite{neonldm}, probing DM masses in the keV range. However, the studied model consisted in DM particles produced by the invisible decay of dark photons generated in the core of a nuclear reactor, which only enabled the search for DM masses up to 1 MeV. Some studies have interpreted the long-lasting DAMA/LIBRA modulation results \cite{damamodold,damamod} in relation to the detection of DM scattering on electrons \cite{damawimpelectronold,damawimpelecmain}. These studies have determined the values for DM mass and cross section with electrons related to the best-fit to DAMA modulation data. While light DM-electron scattering alone cannot fully account for the modulated signal across the entire DAMA/LIBRA energy spectrum, it has been proposed as a potential contributor, particularly in the low-energy region near 1 keV. Also, especially with the threshold reduction of DAMA/LIBRA's crystals to 0.75 keV, this model can only partially explain DAMA's modulation results, with good fitting near the 1 keV energy range, but poor accordance for energies up to 6 keV. However, due to the lack of searches for the DM-electron scattering model in NaI(Tl) detectors, investigating it with COSINE-100 can be highly useful for confirming and expanding the restrictions on the DM-electron cross section.

This work presents the results of the search for events generated by sub-GeV DM scattering on electrons in the COSINE-100 experiment. Excess events over the expected background were searched in a data-set of 2.82 years, using 61.3 kg of NaI(Tl), totalizing an exposure of 172.9 kg-year. A reduced analysis threshold of 8 photoelectrons, corresponding to an energy of 0.7 keV, has been achieved in COSINE-100 crystals \cite{cosineeventselection}, thereby significantly improving the sensitivity for the detection of sub-GeV DM scattering on electrons. The presented study addresses two traditional cases of the DM-electron scattering mediated by a vector boson: a mediator that is much heavier than the electron, resulting in a contact interaction, and a mediator much lighter than the electron, resulting in a long-range interaction.

This paper is structured as follows: in Section \ref{sec:cosine}, the COSINE-100 detector, data acquisition system, and operation are described. In Section \ref{sec:model}, the DM-electron interaction model is detailed, with emphasis on the main relevant expressions for direct detection studies. A brief discussion about an accurate description of the NaI ionization factor appropriate to the energies involved in this analysis is also presented. In Section \ref{sec:data}, the data analysis procedure is reported, describing data treatment and the detector performance at the region of interest (ROI), background components, expected spectra generated by the DM-electron scattering, signal fitting and the resulting constraints for COSINE-100. In Section \ref{sec:sens}, prospects for a more sensitive analysis in COSINE-100 and in its upgraded version COSINE-100U are presented. Finally, a summary highlighting the main results and contributions of this work is provided in Section \ref{sec:concl}.

\section{\textbf{COSINE-100 experiment}}\label{sec:cosine}
The COSINE-100 experiment was installed at the \textit{YangYang} underground laboratory in South Korea, and operated from September 22, 2016, to March 14, 2023. The shortest distance between the surface and the laboratory is 700 m. The main detectors were 8 NaI(Tl) ultrapure crystals, with a total of 106 kg, which were connected in each end to a Hamamatsu R12669SEL photomultiplier tube (PMT). Between the PMTs end faces and the crystals there were 12 mm quartz windows light guides. Oxygen free copper cases encapsulated the crystals, which were immersed in 2200 L of a liquid scintillator (LS) based on linear alkylbenzene, and positioned in a 4 $\times$ 2 array on acrylic tables.

\begin{figure}[!htb]
    \centering
    \includegraphics[scale=0.49, trim = 0mm 0mm 0mm 0mm]{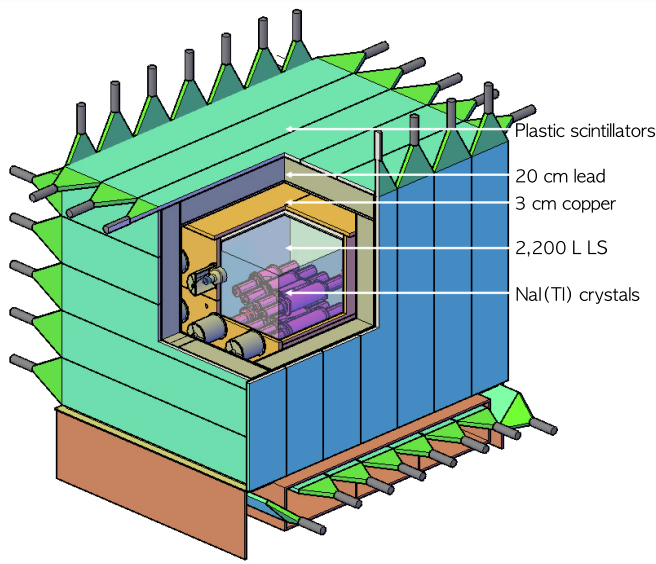}
        \caption[Illustration of the COSINE-100 experiment. The NaI(Tl) crystals arrangement, LS and plastic panels, and passive shields are presented.]{Illustration of the COSINE-100 experiment. The NaI(Tl) crystals arrange, LS and plastic panels, and passive shields are presented.}
    \label{fig:cosine}
\end{figure}

Active and passive vetoes were present in the COSINE-100 experiment. Surrounding the LS, there was a 3 cm thick copper shield and a 20 cm thick layer of low activity lead bricks. In addition, 37 Eljen EJ-200 plastic scintillator panels provided an active veto against cosmic muons. The LS acted not only as a passive and active veto, but was also crucial for measuring the activity of radioisotopes contributing to the background in the crystals. Fig. \ref{fig:cosine} shows a schematic view of the COSINE-100 experiment.

Each PMT coupled to the crystals provided both dynode and anode readouts. The dynode signals are used for analyses in the energy range above 100 keV up to a few MeVs, where saturation effects become relevant. In contrast, the anode signals were employed for low energy analyses, such as the one presented in this work. Even though all 8 crystals are used in analyses above 100 keV, 3 crystals have a poor performance for low energy signals due to low light yield and high noise and background rate. Hence, only data from the remaining 5 crystals are suitable for low energy analysis, and have been used in COSINE-100 WIMP searches \cite{cosinemod,cosinewimpnew}. The total effective mass of these 5 crystals is 61.3 kg. 

Anode signals from the crystals' PMTs were amplified by a factor of 30 and dynode signals amplified by a factor of 100 before processed by 500 mega sample per second (MSPS) FADC modules. As for the active vetoes, signals were read by 62.5 MSPS analog to digital converter (M64ADC) modules. Signals from the plastic scintillators and crystals were in the core of the trigger logic. If the pulse height of a signal in any crystal PMT rises above a threshold of 6 mV, a coincidence gate of 200 ns was generated waiting for another signal above threshold in the other PMT in the same crystal. When the coincidence was observed, the TCB created a global trigger and data from all FADC and M64ADC modules were stored. A global trigger could also be generated if two or more plastic panels observed coincident events. Hence, LS events could only be stored if signals from at least one crystal or two plastic panels were triggered.

Variables from the laboratory environment and detector stability parameters were monitored. Detectors event rate, behavior of parameters used for event selection, and other detector aspects related to its stability were daily monitored by the collaboration. Environment variables were also controlled through the Grafana application \cite{grafana}, as for example the PMTs voltage levels, relative humidity, radon air concentration and the detectors temperature. More details of the COSINE-100 experiment detectors performance, data acquisition system, and environmental monitoring are reported in Refs. \cite{cosineinitperf,cosineliquid,cosinemuon,cosineenv}.

\section{DM-electron scattering model}\label{sec:model}
According to momentum and energy conservations, WIMP-like DM deposits more energy in the target when scattering off particles with comparable mass. Hence, the recent focus on searches for light DM, especially below 10 GeV, has motivated the increasing interest in WIMP-like DM interaction with electrons, since in this mass range, DM particles can produce detectable signals above the threshold of current detectors.

In the following subsection \ref{sec:formulas}, the main concepts and formulas relevant for direct detection searches within the DM-electron scattering model are presented. Subsection \ref{sec:fion} provides a brief discussion of the computation of atomic ionization factors, including the impact of electronic wavefunctions modeling on the expected DM–electron signals and the choice of models appropriate for the analysis presented in this work.
\subsection{Overview and key expressions for direct detection} \label{sec:formulas}
\begin{figure}[!htb]
\centering
\includegraphics[scale=0.95, trim = 60mm 210mm 60mm 45mm]{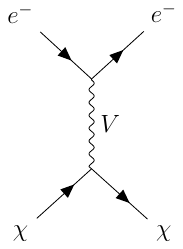}
\caption[Feynman diagram that represents the DM-electron scattering, where $V$ is a vector boson and $\chi$ is the DM particle.]{Feynman diagram that represents the DM-electron scattering, where $V$ is a vector boson and $\chi$ is the DM particle.}
\label{fig:feynman}
\end{figure}

Although the standard DM-nucleon scattering scenario assumes a weak interaction \cite{wimpweak}, there are no restrictions in the models to DM interacting simultaneously with nucleons and electrons, despite a new vector boson is commonly postulated in the latter framework \cite{wimpelectronandnucleon}. The most frequently considered model for the interaction between sub-GeV DM and electrons suggests a hidden sector boson to be the mediator, without restrictions to its spin and parity \cite{damawimpelectronold}. However, in order to simplify the model and accommodate the well motivated scenario of the dark photon \cite{dp}, this boson is supposed to be a vector. In this search, we consider only the scattering between DM and electrons, neglecting the DM-nucleon interaction, in order to directly compare the analysis with reported studies from other experiments. Nevertheless, the typical momentum transfer involved in the scattering is much smaller than the sodium and iodine nuclei masses, implying that the nuclear recoil energy would be negligible. In this scenario, the Feynman diagram representing the scattering is presented in Fig. \ref{fig:feynman}.

Generally, describing the kinematics of the electron ionization by a DM scattering is more complex than the typical DM-nucleon interaction, because the possible values of the momentum transfer in the interaction ($q$) can be as small as O(100 eV) for DM masses in the MeV range and energy depositions of O(1-10 eV), or as high as O(10 MeV) for DM masses in the GeV range and energy depositions of O(1-10 keV) \cite{lagrangianwimpelec,damawimpelectronold}.

Focusing on the direct detection of electron recoils, the velocity averaged cross section for the DM-electron scattering in a detector target can be determined by \cite{xenon10wimpelectron}:

\begin{equation}\label{eq:sec}
    \frac{d\,\langle\sigma_e\,v_\chi\rangle}{dE}=\frac{\overline{\sigma}_e}{2\,m_e}\,\int dv\,\frac{f(v_\chi)}{v_\chi} \int_{q_-}^{q_+} a_0^2\,q\,F_{\chi}^2\,f_{ion}(E,q)\,dq
\end{equation}
\vspace{0.01 cm}

\noindent where $m_e$ is the electron mass, $f(v_\chi)$ and $v_\chi$ are the DM velocity distribution and DM velocity in the Earth frame, respectively, $a_0$ is the Bohr radius, written as $a_0=\frac{1}{m_e\,\alpha}$, $f_{ion}(E,q)$ is the ionization factor, and $q_+$ and $q_-$ are the maximum and minimum momentum transfer, respectively, given by  $q_{\pm}=m_{\chi}v_\chi \pm \sqrt{m^2_{\chi}v_\chi^2-2m_{\chi}E}$, where $m_\chi$ is the DM mass, and $E$ is the energy deposition.

The DM form factor ($F_{\chi}(q)$) and the WIMP-electron scattering cross section (${\overline\sigma}_e$) are the main expressions that contain information about the interaction mediator. These are given by \cite{damawimpelecmain}:

\begin{equation}\label{eq:formfac}
F_{\chi}\,(q)=\frac{\big(\frac{m_{\chi}}{m_e}\big)^2 + \alpha^2}{\big(\frac{m_{\chi}}{m_e}\big)^2 + (\alpha\,a_0\,q)^2}
\end{equation}
\vspace{0.2 cm}
\begin{equation}
\overline\sigma_e=\frac{16\,\pi\,a_0^2\,\alpha^2\,\alpha_{\chi}^2}{\Big(\big(\frac{m_{\chi}}{m_e}\big)^2 + \alpha^2\Big)^2}
\label{exp:xsec}
\end{equation}

\noindent where $\alpha$ is the fine structure constant, $\alpha_{\chi}$ is the DM-electron coupling strength. $\overline\sigma_e$ is the interaction cross section considering a free electron whose scattering with the DM particle transferred a momentum of $a^{-1}_0$ \cite{dmelectroncoupling}. 

Since $\overline\sigma_e$ and $F_\chi(q)$ depend on the mediator mass, studies usually consider two limiting cases, such that a standard scenario is established, and a precise comparison between different results becomes simple. The first case considers a heavy mediator ($m_{\chi}>>\alpha\,m_e$), therefore, $F_{\chi}(q) \longrightarrow 1$. The opposite case considers a light mediator ($m_{\chi}<<\alpha\,m_e$), therefore, $F_{\chi}(q) \longrightarrow \dfrac{1}{(a_0\,q)^2}$.
/
The standard halo model has been assumed in this analysis:

\begin{equation}
    f(v,t) \propto v^2~exp\Big(\frac{-(\vec{v}+\vec{v_E}(t))^2}{v_0^2}\Big)\,\Theta(v-v_{esc})
\end{equation}
\vspace{0.01 cm}

\noindent where $\vec{v}$ is the DM velocity in the galactic frame, $\vec{v_E}(t)$ is the Earth velocity in the galactic frame, $v_0=220$ km/s is the Sun rotation velocity with respect to the galactic center, and $v_{esc}=550$ km/s is the Milky Way escape velocity. $\vec{v_E}(t)$ depends on the relative velocity of Earth with respect to the Sun, and the peculiar velocity of the solar system in the galactic frame. This parametrization is in accordance with the procedure described in Ref. \cite{mod}.

The raw event rate in NaI(Tl) can be determined by:

\begin{equation}
    \frac{dR}{dE} = \frac{n^{NaI}_a\,\rho_{DM}}{m_{\chi}}\,\frac{d\,\langle\sigma_e\,v_\chi\rangle}{dE}
\end{equation}
\vspace{0.01 cm}

\noindent where $n^{NaI}_a$ is the atomic number density per kilogram of NaI, and $\rho_{DM}=0.4$ GeV/cm$^3$ is the approximate energy density of CDM at Earth \cite{dmdensity,damawimpelectronold}.

The parameters chosen for the DM velocity distribution and local density are the same as those adopted in Ref. \cite{damawimpelecmain}, enabling a direct comparison with the DAMA/LIBRA constraints. It is important to note that these values differ from those typically assumed in DM–electron scattering searches \cite{dmparameters}. 
The constraints presented in this work are expected to be less than 25\% more stringent than those obtained using the standard parameter choices. Therefore, comparison with results from other direct detection experiments should be made with appropriate care.

\subsection{Atomic ionization factor} \label{sec:fion} 

Different descriptions are considered when determining the $f_{ion}$ parameter in light DM-electron interaction studies, and an accurate description is essential depending on the target and energies under study, as very divergent results can be obtained \cite{newrelionfac}.

The ionization factor describes the electronic structure of the target material, and different models can be considered depending on the deposited energies in the detector. The first studies conducted in xenon and argon detectors used to assume an analytic non-relativistic approach when computing $f_{ion}$, considering Slater-type orbitals for the bound electrons wavefunctions \cite{darksidewimpelectronold,xenon10wimpelectron}. This procedure leads to reasonable approximations when analyzing very low energy events (typically of O(1-10 eV)), in which few electrons are ionized. However, it neglects important phenomena as the non-point like behavior of nuclear charge for the electron wavefunction near the nucleus, and relativistic effects, which are important for higher energy analyses. For the case of this work, since the ROI is in the keV region due to the 0.7 keV analysis threshold of COSINE-100 crystals, a relativistic description of the electronic structure is important for an accurate analysis of the DM-electron scattering \cite{damawimpelectronold}.

For this analysis, sodium and iodine atoms can be considered to be free, as the typical crystal band structure energies are much lower than the keV scale \cite{essignai,exceeddm,damawimpelecmain}. Therefore, including the ionization factor contribution of NaI(Tl) crystal structure would marginally impact the results of this work. In general, when calculating $f_{ion}$ for each atom isolated, it can be determined from the electron wavefunctions of each shell. However, very distinct results are obtained with different modeling of this atomic electronic structure. More details about the analytic models and a discussion and comparison between non-relativistic and relativistic methods of $f_{ion}$ computation are presented in Appendix \ref{sec:appA}.

$f_{ion}$ is dependent on both $E$ and $q$. Typically, for energy depositions in the keV region, there is an important contribution of $f_{ion}$ in the range of $q \gtrsim 1$ MeV to the event rate. For such high momentum transfer, there is a high correlation of $f_{ion}$ with the radial electron wavefunction near the nucleus, where relativistic effects and correct nucleus size modeling are crucial. Hence, this analytical approach reproduces $f_{ion}$ with good accuracy only at small $q$, suitable for very low energy depositions in the 1–100 eV range in materials whose atoms interaction is negligible, such as noble-gas targets. This approximation, however, is not valid for crystalline materials such as Ge, Si, or NaI(Tl).

A more complete model for a keV region analysis should account for relativistic effects. A possible procedure is to numerically solve the Dirac equation, correctly considering the atomic potential for different distances of the electron from the nucleus by using the relativistic Hartree-Fock approximation \cite{damawimpelectronold}. This approach has been used in the DAMA/LIBRA results interpretation under the DM-electron scattering model. Computation of $f_{ion}$ for sodium and iodine atoms, considering them as free, was performed by authors of the DAMA/LIBRA results interpretation study, and are publicly available in \cite{gitionfac}. The electronic structure and methods for calculations are thoroughly discussed in \cite{damawimpelectronold,damawimpelecmain}. These results are used for the DM-electron interaction spectra generation in the analysis presented in this work.
 
\begin{figure}[!htb]
    \centering
    \includegraphics[scale=0.354, trim = 5mm 5mm 0mm 0mm]{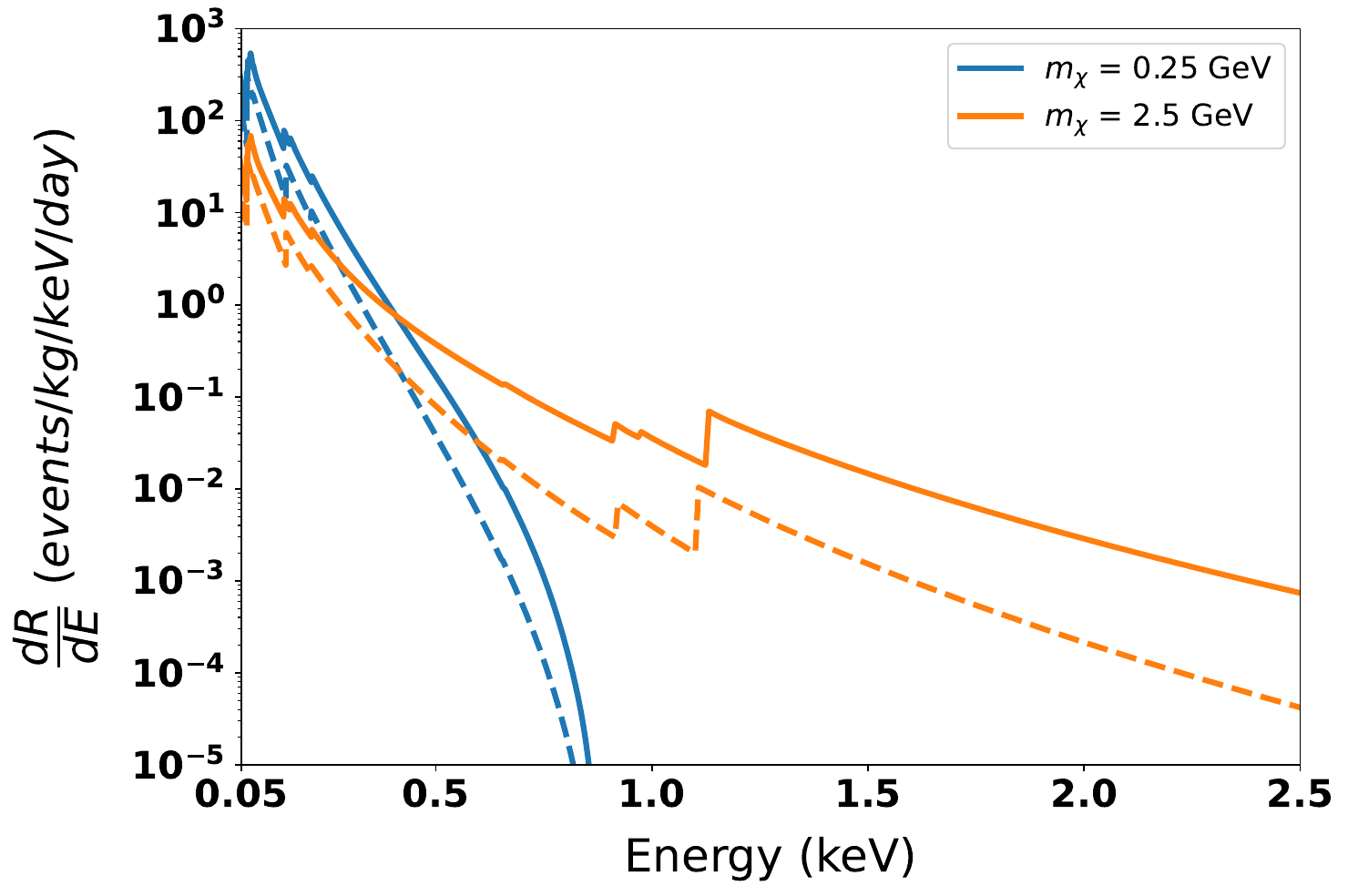}
    \caption[Computed spectra in NaI using the non-relativistic and relativistic methods, considering a heavy mediator. In the blue curves, $m_\chi=1$ GeV is considered, and in the orange curves, $m_\chi=10$ GeV is considered. Solid lines represent the relativistic model, while dashed lines represent the non-relativistic approach.]{Computed spectra in NaI using the analytic non-relativistic and relativistic methods, considering a heavy mediator. In the blue curves, $m_\chi=1$ GeV is considered, and in the orange curves, $m_\chi=10$ GeV is considered. Solid lines represent the computation of spectra with the relativistic approach for $f_{ion}$ calculation, while dashed lines represent spectra with the non-relativistic approach for $f_{ion}$ calculation. The kinks observed between 0.05 keV and 0.3 keV and between 0.8 keV and 1.2 keV arise from electron ionization energies associated with different shells of Na and I atoms.}
    \label{fig:comparison}
\end{figure}

A comparison of the expected spectra in NaI obtained using the analytic non-relativistic approach, described in Appendix \ref{sec:appA}, and the relativistic method for obtaining $f_{ion}$ is shown in Fig. \ref{fig:comparison}. A notable difference between both approaches is a slight shift in the ionization energies due to distinct treatments of electronic wave functions. Relativistic effects lead to small corrections to the electron binding energies relative to calculations for the non-relativistic model, as can be seen by comparing the electron binding and ionization energies reported in Refs. \cite{RHFlow,damawimpelectronold}. In Fig. \ref{fig:comparison}, however, the ionization energies are fixed to the values of the relativistic approach to enable a more direct comparison between the two models.

Regardless of the chosen treatment of atomic electrons, the iodine contribution is dominant compared to sodium for the ROI of this analysis. At 0.1 keV, sodium reaches its maximum contribution of 12\% to the total event rate, and decreases down to 0.3\% at 0.7 keV. Due to convolution of the raw spectrum of Fig. \ref{fig:comparison} with the detector energy resolution function, a small part of events below the 0.7 keV threshold can be detected. Hence, both iodine and sodium atomic ionization factors have been considered in this analysis.

Thallium atoms in NaI(Tl) crystals can also be ionized by the DM particle and contribute to the expected event rate. However, even though thallium has roughly 53\% more electrons than iodine atoms, its concentration in COSINE-100 NaI(Tl) crystals is approximately 0.1 mol\% \cite{NaIcosine200,NaIr&d}, which makes its contribution to be insignificant in this analysis. Naive calculations assuming the non-relativistic approach have shown that the maximum thallium contribution is 2\% at 0.9 keV, and therefore it was neglected. 

For lower energy depositions, the raw event rates using both $f_{ion}$ models are closer to each other, since only low and intermediate $q$ values (typically below 1 MeV) are dominant at this energy region. It is important to note that at energy depositions typically below 0.1 keV, the crystal band structure of NaI(Tl) becomes relevant, and the treatment of isolated sodium and iodine atoms adopted here cannot be used \cite{essignai}. In the most relevant energy region for this analysis, near the 0.7 keV threshold, the disagreement reaches approximately one order of magnitude. As a consequence, the analytic approach underestimates the expected event rate, leading to a reduction in sensitivity. Hence, the use of the more accurate relativistic method is essential for describing energy depositions in the keV region.

\section{Data analysis}\label{sec:data}
To search for DM–electron scattering events in the COSINE-100 data, a detailed treatment of the low energy signals, along with the understanding of the crystals' response and background components in the ROI is required.

The following section \ref{sec:analA} describes the selection of low energy COSINE-100 data suitable for the search for the DM–electron interaction. Section \ref{sec:analB} discusses the expected response of the NaI(Tl) crystals to DM electron signals and its impact on this analysis. Section \ref{sec:analC} presents the background model, including a discussion of the main components and uncertainties relevant to the ROI. Section \ref{sec:analD} describes the fit of the DM–electron energy spectra to the COSINE-100 data.

\subsection{Event selection} \label{sec:analA}
Near the 0.7 keV energy threshold, the majority of detected signals are from PMT-induced noise events, which populate both single-hit and multiple-hit crystals spectra. A single-hit event is classified when only one crystal detects two coincident signals in its PMTs, and no other hits are observed in the LS within a 8 $\mu$s window, or other crystals within a 4 $\mu$s window. A multiple-hit event is determined when hits are observed in two or more crystals, or in the LS and in at least one crystal, satisfying the coincidence time windows mentioned above. Due to the low cross section expected for the DM-electron interaction, the probability of a DM particle hitting multiple electrons is negligible. Thus, only single-hit events were considered in this analysis.

With the objective of greatly reducing the number of noise events in the ROI, a deep learning method based on the multilayer perceptron (MLP) neural network was used \cite{mlp}, allowing the reduction of analysis threshold from the previous 1 keV to 0.7 keV, maintaining an excellent event selection efficiency at the threshold level. Pulse shape discrimination likelihood parameters were introduced as inputs to MLP. The waveforms mean time and amplitudes were evaluated to obtain a PSD parameter. The fast Fourier transform of waveforms has also been included in the generation of another PSD parameter based on the waveforms power spectra. Scintillation events from a one month calibration campaign using an external $^{22}$Na source, together with PMT-induced noise events, were used as inputs for the MLP sample training. Pure scintillation signals were selected by requiring multiple-hit events in the crystals. The $^{22}$Na decay produces a 1274 keV gamma ray from the de-excitation of the $^{22}$Ne daughter nucleus, as well as two 511 keV gamma rays from positron–electron annihilation \cite{na22}. Therefore, pure low-energy scintillation signals from Compton scattering in a given crystal could be selected by requiring a coincidence with high-energy events in another crystal. A MLP score based on the likelihood parameters was calculated for each signal, and the event selection cut determined following the criteria that less than 1\% of the selected events were noise signals. The efficiency of this noise and scintillation event selection is discussed in details in the most recent COSINE-100 WIMP-nucleon interaction search \cite{cosinewimpnew} and in the study of software techniques for lowering COSINE-100 detector threshold  \cite{cosineeventselection}. In addition, the trigger efficiency is verified using a waveform simulation of the PMTs signals \cite{wavesim} and is found to be consistent with unity at the current 8 photoelectrons threshold.

\begin{figure}[!htb]
\begin{center}
   \includegraphics[scale=0.49, trim = 4mm 2mm 0mm 0mm]{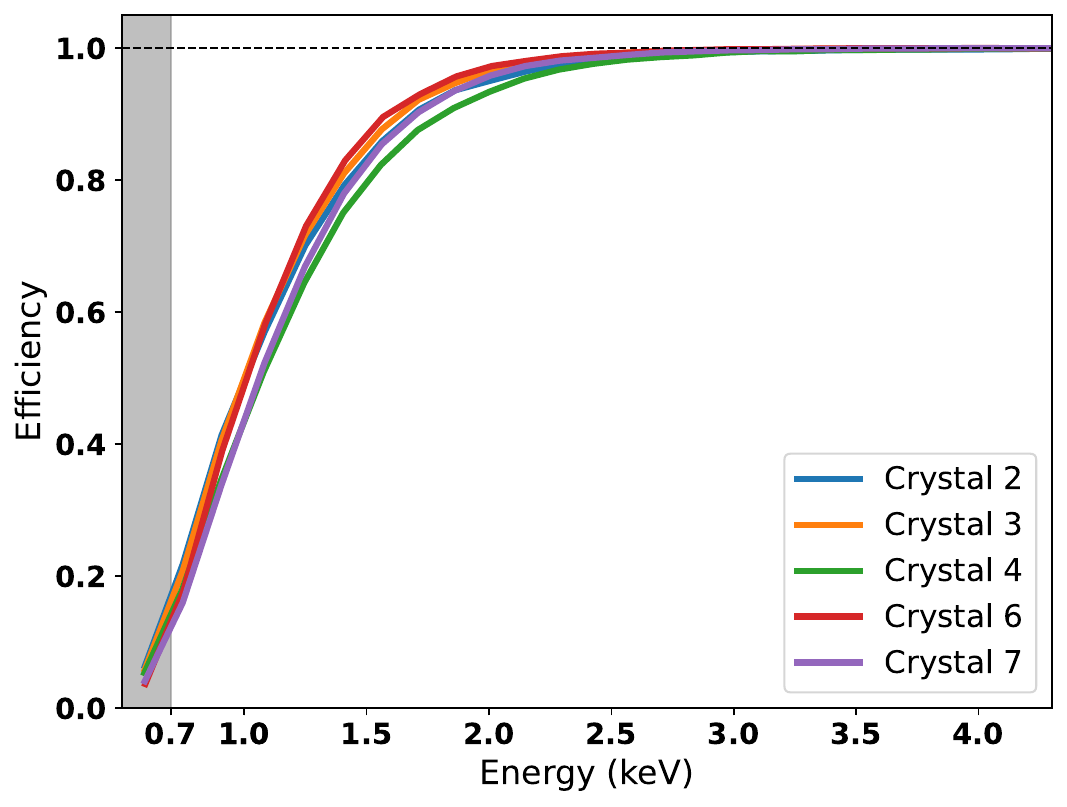}
\caption{Event selection efficiencies as a function of energy for the 5 crystals used in this analysis. The gray shaded region indicates the energy range below the 8 photoelectron analysis threshold. The dashed black line denotes 100\% efficiency.}
\label{fig:eff}
\end{center}
\end{figure}

Other event selections were also applied to data in order to remove possible muon-induced events, as well as noise events originated from electronics. Regarding the latter, only events that generated at least two clusters in the waveform, and whose first trigger pulse was separated by at least 2 $\mu$s from the beginning of the event recording were selected. Events induced by muons in the crystals were avoided by selecting only events that were not coincident with muon signals in the plastic scintillators within a 30 ms window. This is especially important, since muons might induce neutron events that deposit energies in the ROI of this analysis. The event selection efficiency considered in this analysis is shown as a function of energy in Fig. \ref{fig:eff}.

\subsection{COSINE-100 crystals response and DM-electron scattering spectra} \label{sec:analB}
In order to generate the expected spectra in COSINE-100 crystals for the DM-electron scattering model, it is essential to properly consider the detector's energy resolution and the non-linear proportionality between the measured charge in the PMTs and deposited energy at low photoelectrons level. As a consequence of the small energy deposition of DM on electrons, the number of events should be higher below the 8 photoelectrons threshold. The detector performance in this region has to be rigorously studied, since these events may induce signals exceeding the threshold due to the energy resolution effect. 

The crystals resolution and non-proportionality at very low energies were carefully examined by data from calibrations using the 0.87 keV X-ray emission of the $^{22}$Na isotope and the 3.2 keV X-ray peak of the $^{40}$K isotope, both present as internal contaminants of the crystals \cite{cosinenpr}. The detector behavior has been reproduced down to one photoelectron level in the PMTs waveform simulation \cite{wavesim}. Accordingly to these studies, the following scaled Poisson distribution describes the crystals energy resolution for few photoelectrons yield:

\begin{equation}\label{eq:poisson}
S(q_{raw},q_{vis},s)=\frac{({q_{raw}/s})^{q_{vis}/s}\,e^{-q_{raw}/s}}{\Gamma(q_{vis}/s+1)}
\end{equation}
\vspace{0.01 cm}
    
\noindent where $q_{raw}$ is the deposited charge in the detector, $q_{vis}$ is the visible charge, and the $s$ parameter is determined from calibration \cite{cosinenpr}.

\begin{figure}[!htb]
\begin{center}
   \includegraphics[scale=0.303, trim = 5mm 0mm 0mm 0mm]{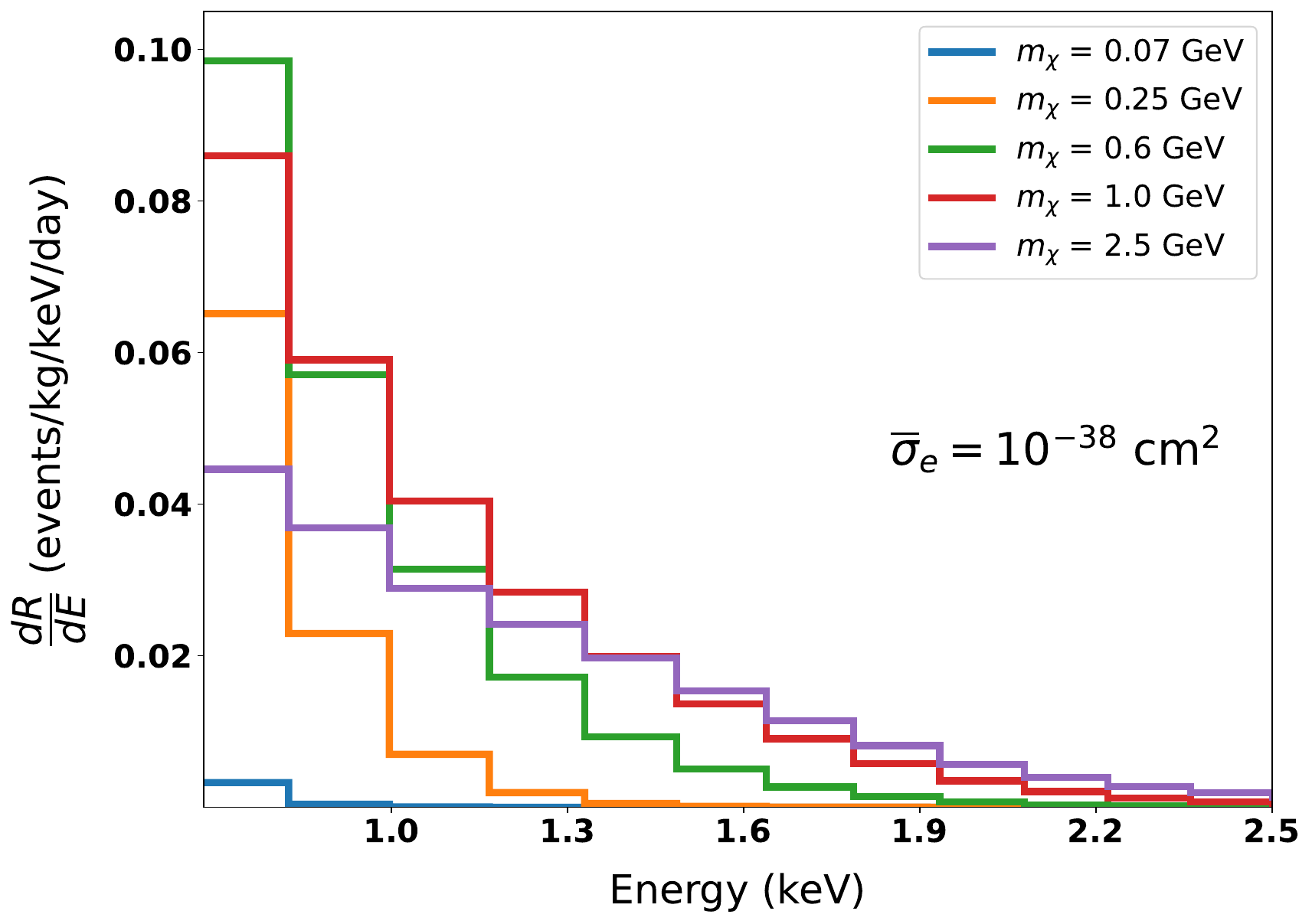}
\caption{Expected event rate in COSINE-100 NaI(Tl) detectors considering a heavy mediator and 5 different DM masses. The curves represent an average over the 5 analyzed crystals. $\overline{\sigma}_e$ is fixed at $10^{-38}$ cm$^2$.}
\label{fig:spectra}
\end{center}
\end{figure}

The generated visible spectra in COSINE-100 NaI(Tl) crystals, corrected by event selection efficiency, detectors' non-proportionality and energy resolution are presented in Fig. \ref{fig:spectra}, assuming the heavy mediator scenario. Most signals are expected to be detected with energies near the 0.7 keV threshold for all DM masses, which makes the ROI of this analysis to be at very low energy depositions. Hence, we considered only data up to 100 photoelectrons (equivalent to approximately 7.8 keV) in the data fit. Also, due to the discussed detector response, the expected number of events is highest for $m_\chi\approx$ 0.6 GeV when considering the heavy mediator case, and for $m_\chi\approx$ 0.2 GeV when considering the light mediator case. Therefore, the COSINE-100 sensitivity is expected to be higher near these DM masses. In addition, for lower DM masses, the expected event rate in the first energy bin is several times larger than that in the higher energy bins. As can be observed in Fig. \ref{fig:comparison}, light DM particles in the halo do not have sufficient kinetic energy to deposit keV scale energies in the detector, resulting in most events occurring below the COSINE-100 threshold. Nevertheless, due to the energy resolution of the detector, a fraction of these events can be reconstructed above 0.7 keV, while retaining a quickly falling spectral shape. Consequently, an accurate understanding of the background near threshold is crucial for enhancing the sensitivity of the analysis to DM particles in the MeV mass range.

\subsection{Background components}\label{sec:analC}
The activities of each component that constitute the COSINE-100 crystals background are known from a detailed detector simulation based on \textit{Geant}-4 \cite{geant}. The detector exposure time of the data-set used for the background modeling is identical to that considered in this analysis. The activity of each isotope is determined through a combined fit to single-hit events in the energy range from 6 keV to 4000 keV and to multiple-hit events from 0.7 keV to 4000 keV. Single-hit events in the [0.7–6] keV region are excluded from the background Monte Carlo (MC) fit in order to preserve blindness to potential low energy DM signals, such as the expected from DM–electron scattering \cite{set3background}.

\begin{figure}[!htb]
    \centering
    \includegraphics[scale=0.317, trim = 8mm 0mm 4mm 0mm]{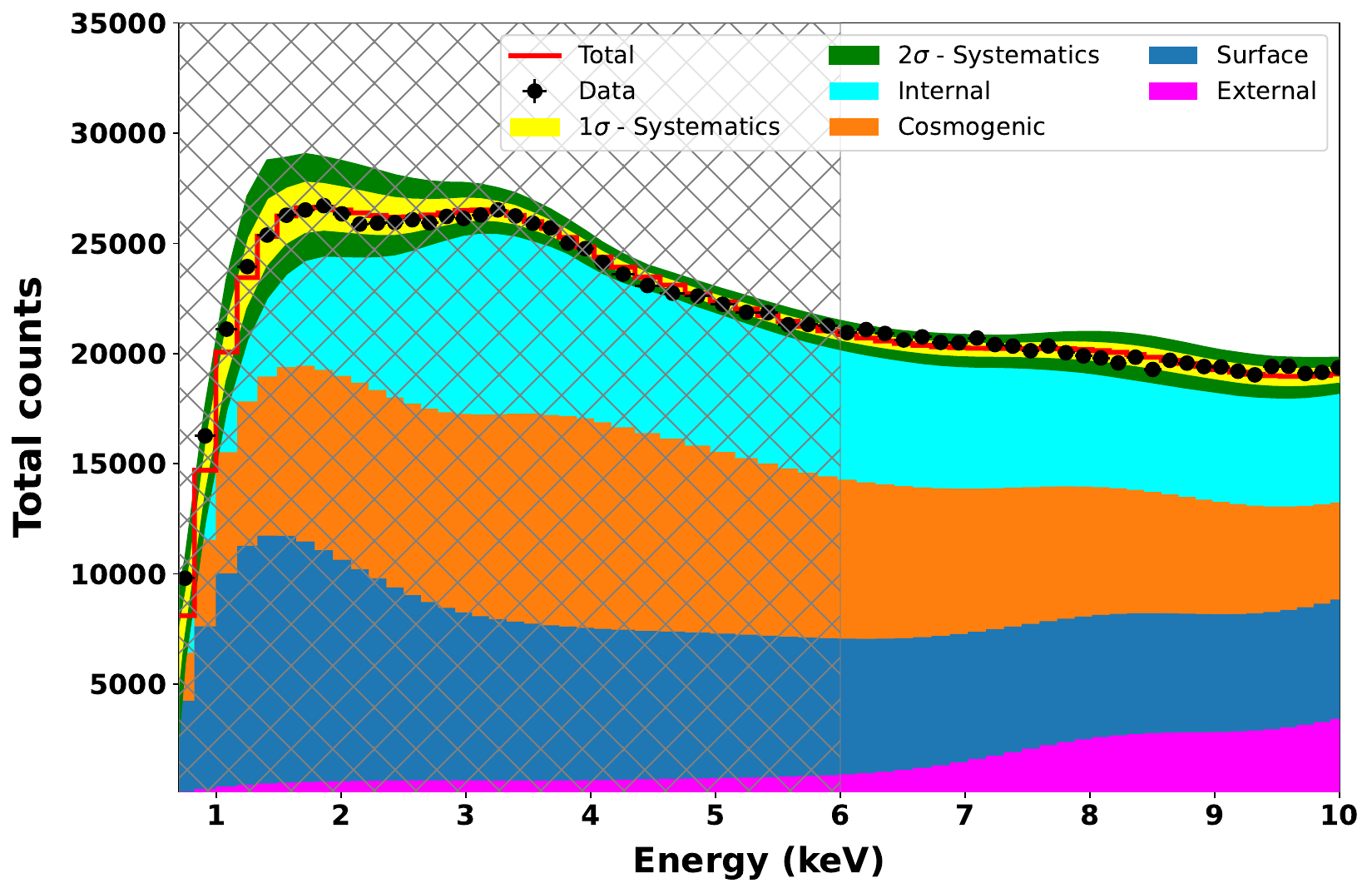}
        \caption{Measured data for the 2.82 years data-set used in this analysis (black points), and the crystals background model near the ROI (red histogram). Data and background for all analyzed crystals are summed. Background components are labeled into 4 groups according to their origin. The yellow and green bands represent the 1$\sigma$ and 2$\sigma$ of the dominant sources of systematics, which constitute the leading sources of uncertainty in the ROI. The hatched region denotes the energy range over which the background model, defined using data above 6 keV, is extrapolated.}
    \label{fig:bkg}
\end{figure}

For energies near the 0.7 keV threshold, the dominant background components are secondary radiations from $\beta$ decays of $^{3}$H and Compton scattered $\gamma$s and $\beta$ decays of $^{210}$Pb, both internal sources in NaI(Tl) crystals, which represent around 75\% of all events detected up to 2 keV. Fig. \ref{fig:bkg} presents the single-hit background model of COSINE-100 crystals considered in this analysis.  

Systematic uncertainties presented in the background model have different sources, with the most relevant at the ROI being uncertainties in the: MLP event selection efficiencies; determination of the precise location of background sources in the crystals' PMTs; evaluation of the parameters that describe the crystals' energy resolution distribution of Equation \ref{eq:poisson}; measurement of the activity of each background component; and depth of the surface $^{210}$Pb in the crystals. Table \ref{tab:table} summarizes the contribution of each systematics component in three energy intervals within the ROI. These contributions were calculated by first evaluating, in each energy bin of the interval, the ratio between its 1$\sigma$ uncertainty and the total 1$\sigma$ systematic uncertainty, and then averaging this ratio over all bins in the corresponding energy interval.

\begin{table*}[!t]
\caption{Contribution of each systematic error component of the background model in the ROI}
\label{tab:table}
\setlength{\tabcolsep}{12pt}
\centering 
\renewcommand{\arraystretch}{1.1} 
\begin{tabular}{l c c c}
\hline \hline 
\textbf{Systematics component} & \multicolumn{3}{c}{\textbf{Contribution ratio (\%)}} \\
\hline 
& \textbf{[0.7 - 1.5] keV} & \textbf{[1.5 - 4] keV} & \textbf{[4 - 7.8] keV} \\
\cline{2-4}
Event selection efficiency       & 67.59 & 12.52 & 11.71 \\
PMT background sources location  & 4.03  & 8.09  & 21.31 \\
Energy resolution parameters     & 0.23  & 2.48  & 6.34  \\
Activity of background components & 2.18  & 11.57 & 28.23 \\
Depth of surface $^{210}$Pb      & 25.98 & 65.34 & 32.40 \\
\hline \hline 
\end{tabular}
\end{table*}

The background model used in this analysis is the same described in the recent COSINE-100 WIMP search \cite{cosinewimpnew}, with the only difference being that a scaled Poisson distribution is considered instead of a Gaussian distribution when correcting the spectra of background components by the energy resolution effect. This is important to match the energy resolution correction applied to the DM-electron interaction signals with the correction of background components spectra. The main different results to the final background model are generated by radioisotopes that generate few photoelectrons, since the shape of the scaled Poisson function differs from the Gaussian function at very low energy depositions. Thus, $^{22}$Na, $^{109}$Cd, and $^{113}$Sn cosmogenic isotopes are the components which differed the most with the modification of the energy resolution function, due to their emission of X-rays and Auger electrons near the 0.7 keV threshold. Nevertheless, when compared to the official COSINE-100 crystals background model \cite{set3background}, the change in the energy resolution distribution marginally increased the number of events in the background model spectrum, up to a maximum of 2\% in the first energy bin.

\subsection{Spectral fit}\label{sec:analD}
A Bayesian analysis approach was used in the spectra fit to the 2.82 years data-set process. The posterior distribution for the DM-electron signal was obtained from a likelihood function based on Poisson probabilities, determined by a Markov Chain Monte Carlo (MCMC) based algorithm. This analysis method is the same adopted in other DM searches in COSINE-100 \cite{cosineboosted,cosineinel,cosinesuperwimp,cosinesolaraxion}, and is thoroughly described in Ref. \cite{cosinewimp}.

\begin{figure}[!htb]
    \centering
    \includegraphics[scale=0.32, trim = 8mm 0mm 0mm 0mm]{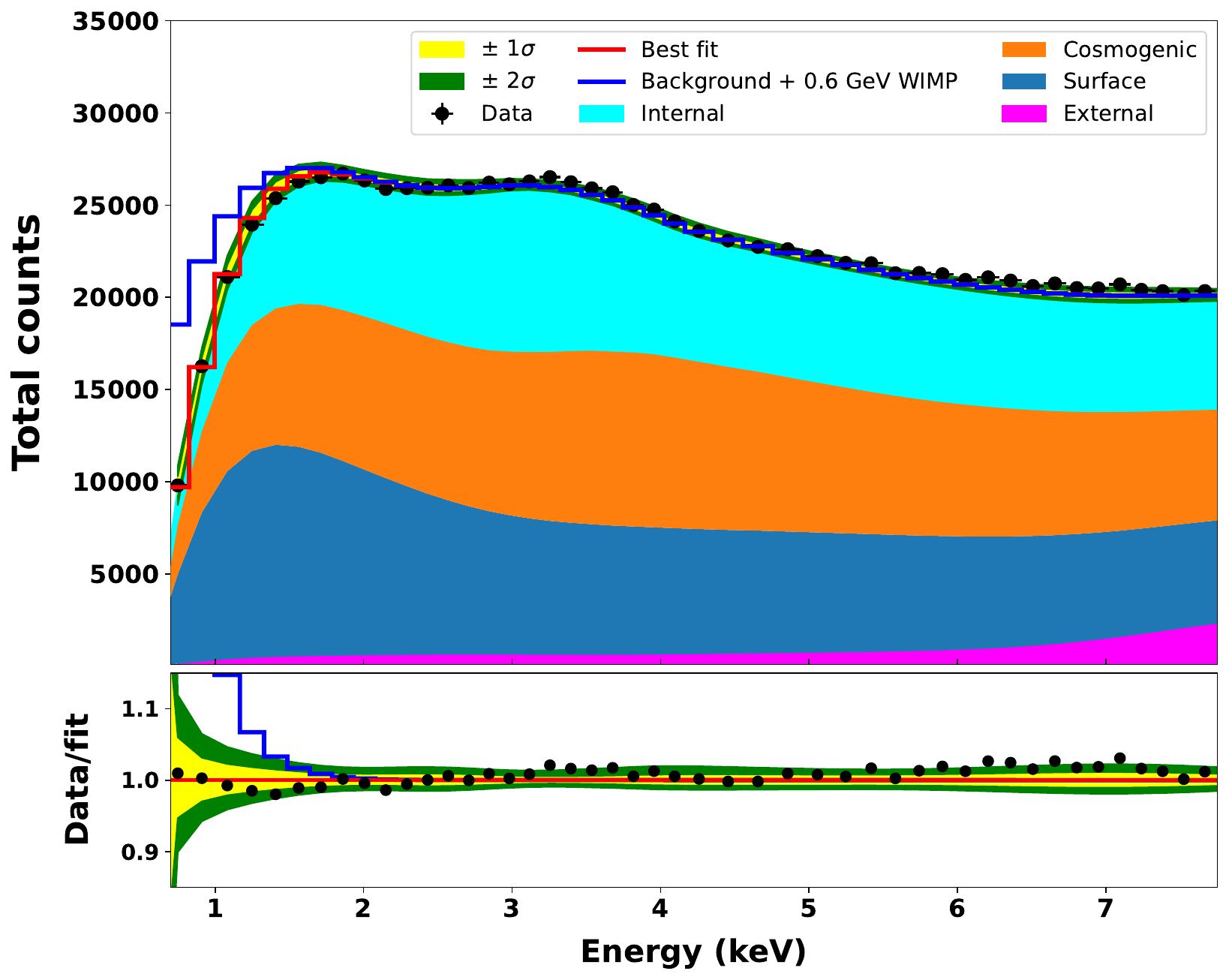}
    \caption{COSINE-100 data fit results for $m_\chi=0.6$ GeV, assuming the heavy mediator scenario. Background components, separated into their different origins, are presented. Yellow and green bands represent the 1$\sigma$ and 2$\sigma$ fit uncertainties, respectively. The expected measured data for the DM-electron cross section of $\overline{\sigma}_e=9.2 \times 10^{-38}$ cm$^2$ that best fits DAMA/LIBRA modulation, presented in Ref. \cite{damawimpelecmain}, is also shown in the blue histogram. The larger systematic uncertainty near the threshold is due to the increasing of systematics associated with the event selection efficiency.}
    \label{fig:fit}
\end{figure}

The fit was performed for several DM masses, ranging from 30 MeV to 5 GeV. The mean background activities and statistical uncertainties, determined from each crystal background model \cite{set3background}, were used to produce the Gaussian priors. Systematic uncertainties were included as nuisance parameters also with Gaussian priors. Each crystal background is allowed to independently fluctuate in the simultaneous fit performed, but the $\overline{\sigma}_e$ distribution is marginalized over all 5 crystals. No evidence of excess events consistent with expected spectra from both the light and heavy mediator models was found. The resulting fit for a 0.6 GeV DM and assuming a heavy mediator is presented in Fig. \ref{fig:fit}, along with the expected observed data for $\overline{\sigma}_e=9.2 \times 10^{-38}$ cm$^2$.  

\begin{figure}[!htb]
    \centering
    \includegraphics[scale=0.465, trim = 8mm 0mm 5mm 0mm]{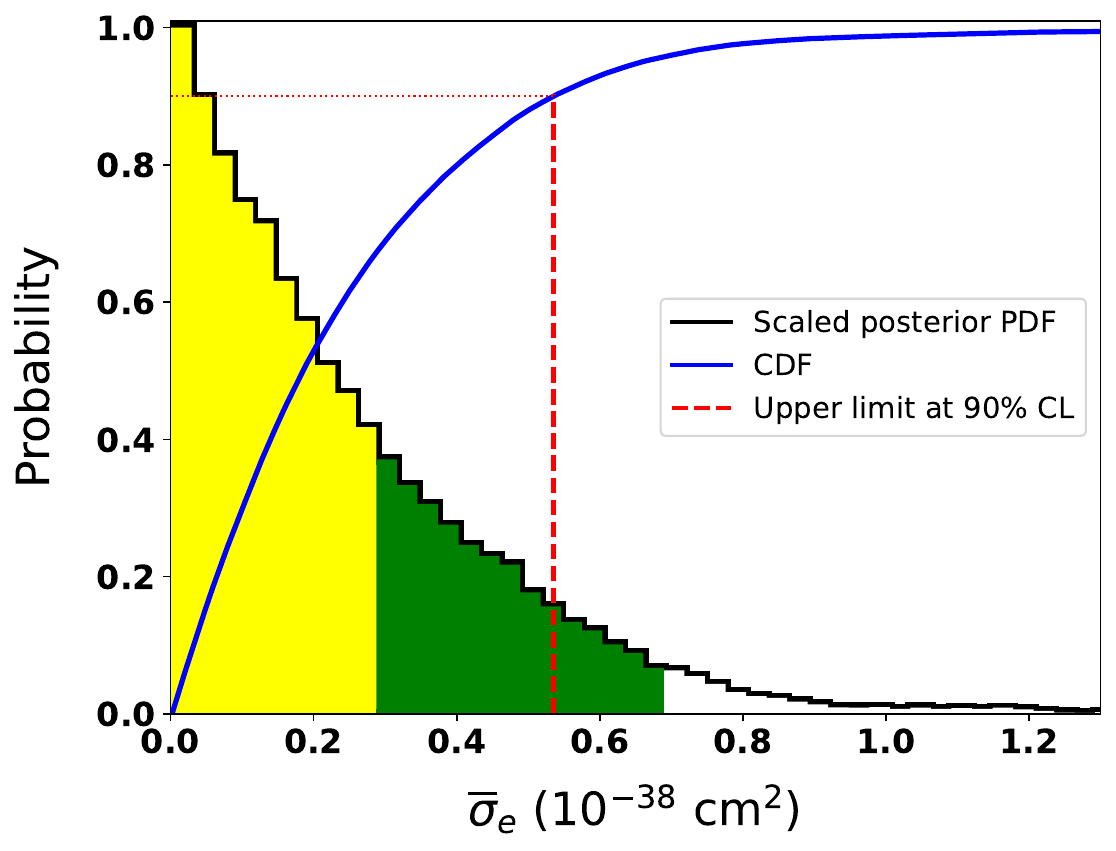}
    \caption{Normalized posterior distribution for $\overline{\sigma}_e$, considering a $m_\chi=0.6$ GeV DM particle and a heavy mediator. Yellow and green bands represent the 1$\sigma$ and 2$\sigma$ CL probabilities, respectively. The blue curve shows the cumulative distribution function (CDF).}
    \label{fig:posterior}
\end{figure}

After marginalization, the resulting posterior probability distribution function (PDF) for the DM-electron signal strength is used to set a 90\% confidence level (CL) upper limits on $\overline{\sigma}_e$. Fig. \ref{fig:posterior} shows the posterior distribution for the 0.6 GeV DM data fit.

\begin{figure*}[!htb]
    \centering
    \includegraphics[scale=0.3475, trim = 11mm 0mm 0mm 0mm]{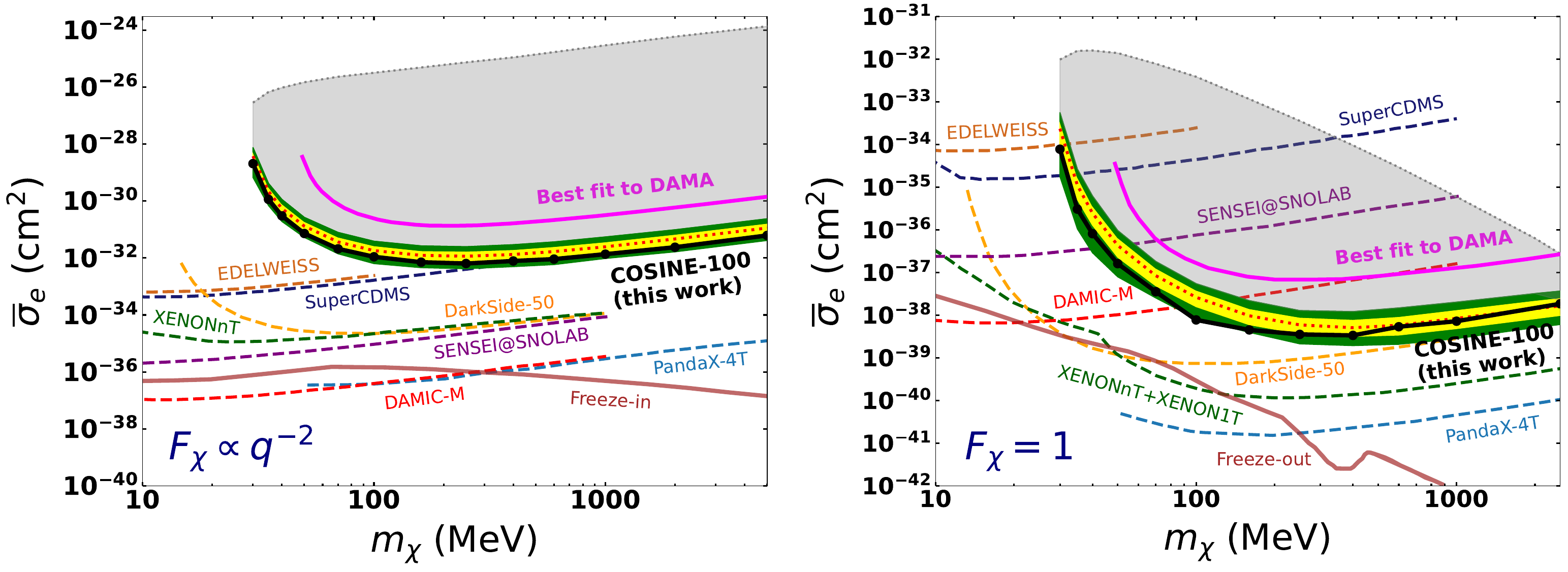}
    \caption{Upper limits for $\overline{\sigma}_e$ considering the light mediator model (left) and heavy mediator model (right). The dotted red lines and the yellow and green areas represent the median 90 \% CL expected sensitivity and its 1$\sigma$ and 2$\sigma$ bands, respectively. The magenta lines are the best fit to DAMA/LIBRA modulation results, presented in Ref. \cite{damawimpelecmain}. Constraints from the PandaX-4T \cite{pandaxwimpelectron}, SENSEI \cite{senseiwimpelectron}, DarkSide-50 \cite{darksidewimpelectron}, XENON1T and XENONnT \cite{xenon1twimpelectronreviewer,xenonntwimpelectron}, DAMIC-M \cite{damicmwimpelectron}, EDELWEISS \cite{edelweisswimpelectron}, and SuperCDMS \cite{supercdmswimpelectron} experiments are also shown. For the heavy mediator case, constraints from XENON1T and XENONnT are combined and shown as a continuous limit. The brown shaded regions show required parameters from models which obtain the observed DM relic abundance by freeze-in (light mediator) or freeze-out (heavy mediator) mechanisms \cite{wimpelectronandnucleon}. The gray shaded region indicates the parameter space excluded by COSINE-100 constraints. DM particles with parameters above the dotted gray lines are unable to reach Y2L due to Earth shielding effects, which was evaluated using the Verne 2.1 code \cite{vernegit,verneheavy,vernelight} (see Appendix \ref{sec:appB} for details). For the heavy-mediator case, only DM masses up to 2.5 GeV are shown, since a non-negligible fraction of particles is expected to be stopped before reaching the detector. Note that the parameters of the DM velocity distribution and local energy density differ from those used for the dashed line constraints (see the discussion in Sec. \ref{sec:formulas}).}
    \label{fig:limits}
\end{figure*}

Simulated data samples based on the background model were produced in order to test the analysis sensitivity. Based on the Gaussian priors of each crystal background component and systematic uncertainties, 1000 pseudo-data samples were generated assuming a background-only hypothesis. The same fitting procedure of the physics data was used, and 90\% CL upper limits were set for each sample. The resulting upper limits distribution for each DM mass was used to perform calculations of the mean sensitivity and its 1$\sigma$ and 2$\sigma$ uncertainty bands. Fig. \ref{fig:limits} presents the limits set for light and heavy mediator models.

For lighter DM masses, in both light and heavy mediator scenarios, the distance between the set upper limits and the median sensitivity remains almost constant for any DM mass. This effect reflects the small change in the spectra shape for low mass DM particles.

The upper limits set from this COSINE-100 data analysis for a light and heavy mediator scenarios do not explore regions unconstrained by the most sensitive experiments for the DM-electron interaction model, due to their capability of reaching very low threshold (down to a single electron ionization \cite{xenon1twimpelectron}). However, the reported COSINE-100 constraints completely exclude the best fit regions for DAMA/LIBRA modulation results assuming any vector mediator mass, and set the current most stringent upper limits for the DM-electron scattering model for a NaI(Tl) experiment.

\section{\textbf{Prospects for COSINE-100 and COSINE-100U}}\label{sec:sens}
The COSINE-100U experiment will be the upgraded version of COSINE-100 \cite{cosine100u}. It will be located at the \textit{Yemilab} underground laboratory, also in South Korea. The new laboratory is 300 m deeper than \textit{YangYang}, reducing even more the contribution of cosmic rays in the experiment. The main objective of COSINE-100U is to enhance the NaI(Tl) crystals light yield and make all 8 crystals suitable for low energy analysis, reducing the detector threshold, and increasing sensitivity to sub-GeV DM searches.

A new encapsulation method, focused on removing the need for a quartz light guide between the crystals' end faces and PMTs, has been proved to reduce scintillation photon losses and improve light collection \cite{crystalencap1,crystalencap2}. The crystals operation at $\SI{-33}{\celsius}$ also provides an increase in light yield \cite{lowtemp}. Both upgrades will be applied to COSINE-100U experiment, and are expected to enhance the light yield by about 40\%. This is expected to enable a reduction of the analysis threshold from 0.7 keV to approximately 0.2 keV.

\begin{figure*}[!htb]
    \centering
    \includegraphics[scale=0.3457, trim = 11mm 0mm 0mm 0mm]{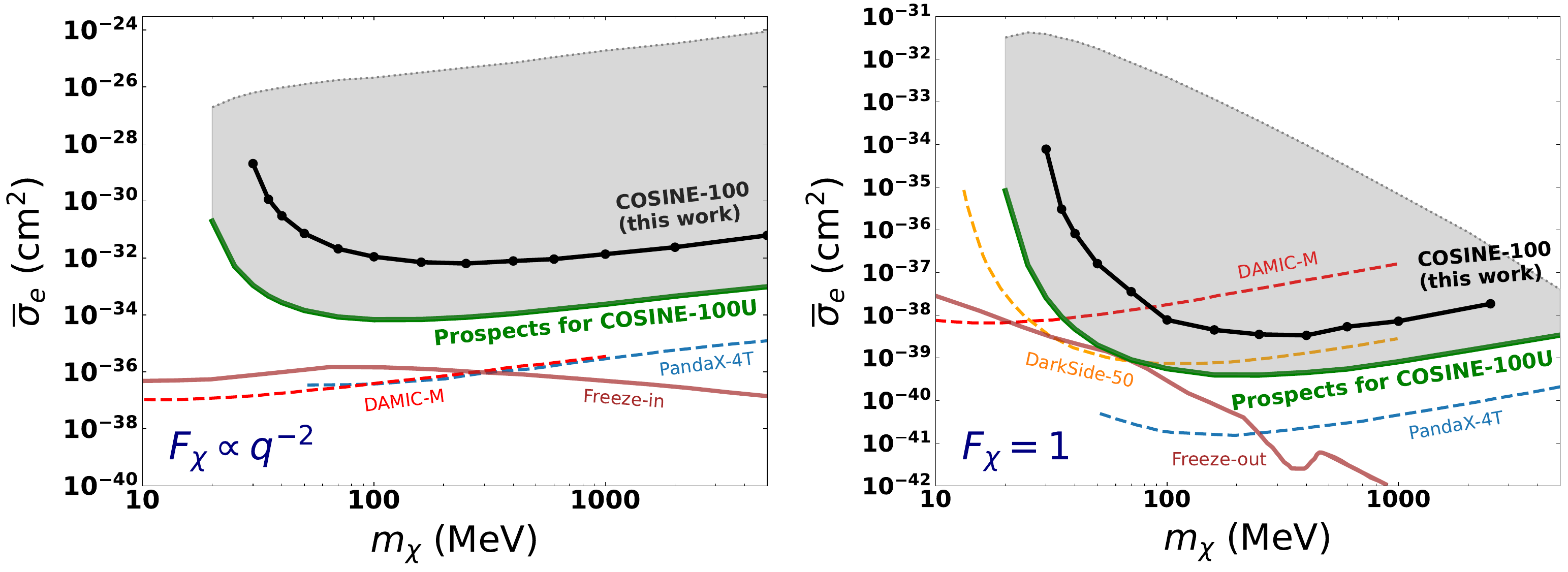}
    \caption{Sensitivity prospects for the COSINE-100U experiment (red) considering the light mediator model (left) and heavy mediator model (right), for 90\% CL. The black data points represent the upper limits set for COSINE-100 in this study. The most stringent constraints from experiments and the models parameters for freeze-in and freeze-out are also shown. The gray shaded region indicates the parameter space expected to be excluded by COSINE-100U. DM particles with parameters above the dotted gray lines are unable to reach Yemilab due to Earth shielding effects (see Appendix \ref{sec:appB} for details). Note that the parameters of the DM velocity distribution and local energy density differ from those used for the dashed line constraints (see the discussion in Sec. \ref{sec:formulas}).}
    \label{fig:cosine100U}
\end{figure*}

Since the expected number of events for the DM-electron interaction in NaI(Tl) is much higher for very low energy depositions, as presented in the spectra of Fig. \ref{fig:spectra}, a threshold reduction of the crystals would remarkably enhance the analysis sensitivity. Fig. \ref{fig:cosine100U} shows an estimated projection of the constraints for COSINE-100U, assuming a 5 photoelectrons threshold (approximately 0.2 keV). The prospects indicate a notable improvement in sensitivity for low mass DM, however, it remains uncompetitive with limits established by other experiments.

An analysis below the current 0.7 keV COSINE-100 threshold, potentially extended down to a one photoelectron level, is currently being investigated. Given that a precise background modeling at such a small energy deposition is unfeasible due to the very high noise rate, an alternative approach in terms of event selection and data analysis will be required. It is proposed to conduct an annual modulation analysis, which is less reliant on the exact determination of background activity, provided that the stability of noise and background rates are ensured. Details of the COSINE-100 crystal data below the 8 photoelectron threshold, as well as the modulation analysis, are left for a future publication. Once the feasibility of this study is confirmed using COSINE-100 data, it can be implemented in COSINE-100U NaI(Tl) crystals to enhance the analysis sensitivity due to the higher light emission and collection efficiency.

To accurately model the DM–electron scattering signals down to the a single photoelectron level, the NaI crystal band structure must be taken into account, and treating the Na and I atoms as isolated is no longer a valid approximation, as discussed in Section \ref{sec:fion}. Once the information about electronic wave functions of NaI crystal is obtained, open source codes such as \textit{QEdark} \cite{wimpelectronandnucleon} and \textit{EXCEED-DM} \cite{exceeddm} can be used to compute the expected DM–electron energy spectra.

Sensitivity studies on the behavior of NaI at low energy depositions have shown promising prospects of this material compared to silicon, xenon, and argon, which constitute the most sensitive experiments to light DM–electron interactions \cite{thesisionfac,essignai}. The NaI energy band gap is typically calculated to be 5.9 eV \cite{essignai}, lying between that of silicon (1.11 eV \cite{qcdark}) and liquid xenon ($\sim\,$9.22 eV \cite{xenonbandgap}). Particularly for the heavy mediator scenario, if an analysis at a one photoelectron level can be achieved, the most stringent sensitivity is expected to lie in the DM mass range of 10–50 MeV \cite{cosinusnai}, where cross sections below $6 \times 10^{-39}$ cm$^2$ remain unconstrained by current experiments.

\section{Conclusion}\label{sec:concl}
We reported the search for sub-GeV DM scattering on electrons using a 2.82 years data-set of the COSINE-100 experiment NaI(Tl) crystals. We found no excess events consistent with the expected signals for the DM-electron interaction assuming a heavy or a light vector boson mediator. 90\% confidence level upper limits were set on the DM-electron scattering cross section. The most stringent constraints exclude cross sections above 6.4 $\times$ 10$^{-33}$ cm$^2$ for a 0.25 GeV DM, assuming a light mediator, and above 3.4 $\times$ 10$^{-37}$ cm$^2$ for a 0.4 GeV DM, assuming a heavy mediator. These results represent the most stringent exclusion limits on the DM-electron scattering model for a NaI(Tl) target to date, and fully cover the parameters attributed to DAMA/LIBRA modulation results interpreted within this framework. We propose a more sensitive study using COSINE-100 NaI(Tl) crystals, aimed at searching for annual modulation signals induced by DM electron interactions in few generated photoelectrons in the PMTs. This study could potentially allow the search for MeV DM particles in NaI(Tl), and will be the subject of a future report.

\acknowledgments
We thank the Korea Hydro and Nuclear Power (KHNP) Company for providing underground laboratory space at Yangyang and the IBS Research Solution Center (RSC) for providing high performance computing resources. We are also very grateful for Ashlee R. Caddell and Benjamin M. Roberts for the helpful discussions and for performing the customized computation of the ionization factors for sodium and iodine. This work is supported by: the Institute for Basic Science (IBS) under project code IBS-R016-A1,  NRF-2021R1A2C3010989, NRF-2021R1A2C1013761, RS-2024-00356960, RS-2025-25442707 and RS-2025-16064659, Republic of Korea; NSF Grants No. PHY-1913742, United States; STFC Grant ST/N000277/1 and ST/K001337/1, United Kingdom;
Grant No. 2021/06743-1, 2022/12002-7, 2022/13293-5 and 2025/01639-2 FAPESP, CAPES Finance Code 001, CNPq 304658/2023-5, Brazil;
LPDP Grant under WCU Equity 2024, Indonesia.

\appendix
\section{Non-relativistic $f_{ion}$ calculation, and comparison with relativistic model}\label{sec:appA}
The analytic calculation method for $f_{ion}$ considered in the non-relativistic spectra presented in Fig. \ref{fig:comparison} is the same used in the studies reported by DarkSide-50 collaboration \cite{darksidewimpelectronold,darksidewimpelectron}. The Roothan-Hartree-Fock (RHF) wavefunctions represent the radial part of bound electrons. These are written as a linear combination of hydrogen-like wavefunctions (Slater-type orbitals) \cite{RHFlow,RHFhigh}.

For ionized electrons, the radial wavefunctions are obtained from the solution to the Schr\"{o}dinger equation with the Coulomb potential $-Z^{nl}_{eff}/r$, where $Z^{nl}_{eff}$ is the effective charge observed by the scattered electron, and $n$ and $l$ indicate the electron shell. This charge is determined in order to reproduce the binding energy of the electron, assuming only the effect of a Coulomb potential.

Near the nucleus (electron wavefunction for $r<<R$, where $R$ is the atom radius), $Z^{nl}_{eff}$ is expected to tend to $Z$, since only the nuclear charge becomes relevant. The opposite happens far from the nucleus (electron wavefunction for $r>>R$), where $Z^{nl}_{eff}$ should tend to 1, as the total charge of the ionized atom becomes unity. This process is not accounted for when defining a fixed value for $Z^{nl}_{eff}$ from the Coulomb potential as described.

It is possible to insert into the analytic model an effective charge dependent on the distance between the electron and the nucleus, derived from the ``frozen core approximation'' \cite{zeff}. In this formulation, $Z^{nl}_{eff}(r)$ can be written by:

\begin{equation}\label{eq:zeffr}
    Z^{nl}_{eff}(r)=Z+\nu_{nl}(r)-\sum_b (2 j_b+1)\,\nu_b(r)
\end{equation}
\vspace{0.01 cm}

\noindent where the sum over $b$ is over all bound electrons, and $2 j_b+1$ is the number of bound electrons in shell $b=l$. $\nu(r)$ is the potential of a single electron, which is obtained from the radial components of electron wavefunctions of each shell by:

\begin{equation}
    \nu_{nl}(r)=\int_0^\infty dr'\,r'^2\;\frac{R^2_{nl}(r)}{r_>}
\end{equation}
\vspace{0.01 cm}

\noindent where $r$ is the distance between the electron and the nuclear center, $r'$ is an integration variable, and $r_>$ is the larger of $r$ and $r'$.

\begin{figure}[!htb]
    \centering
    \includegraphics[scale=0.415, trim = 5mm 4mm 0mm 4mm]{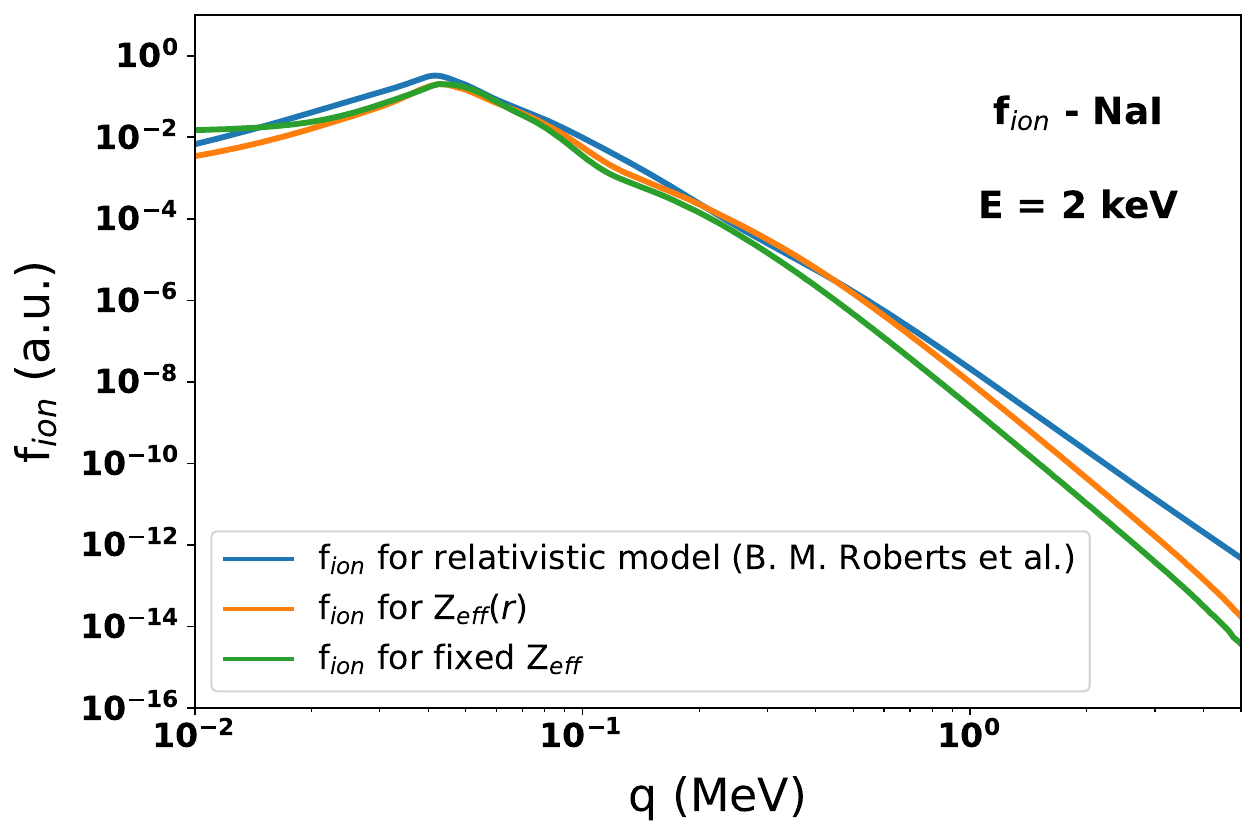}
    \caption{Comparison between different methods for performing the $f_{ion}$ calculation for NaI, considering a 2 keV energy deposition. The $f_{ion}$ calculated by the relativistic method \cite{newrelionfac,gitionfac}, by the $Z^{nl}_{eff}$ dependent on $r$ model, and by the fixed $Z^{nl}_{eff}$ model are presented by the blue, orange and green curves, respectively.}
    \label{fig:fioncomparison}
\end{figure}

In contrast to the case of a fixed $Z^{nl}_{eff}$, the effective charge of the ionized electron varies with $r$. Furthermore, there is no significant difference in the dependence of $Z^{nl}_{eff}$ on $r$ for electrons from any shell, given that the effective charge is barely sensitive to the ground state electron shell. Fig. \ref{fig:fioncomparison} shows the comparison between $f_{ion}$ calculated according to the analytic method assuming a fixed $Z^{nl}_{eff}$, a $Z^{nl}_{eff}$(r) (dependent on $r$), and according to the relativistic method. Calculations were performed for Na and I atoms considering a 2 keV energy deposition. For this energy deposition, the iodine electrons of the 1s, 2s, and 2p shells do not contribute to $f_{ion}$, since their binding energies are higher than the deposited energy.

Despite resulting in a reasonable description of $f_{ion}$ for intermediate values of $q$ (generally in the 100 keV to 1 MeV range), this non-relativistic approach fails to accurately describe the electron scattering behavior for small and high momentum transfers.

Updating the non-relativistic model with a $Z^{nl}_{eff}$ dependent on $r$ is able to make $f_{ion}$ more compatible with the relativistic approach, particularly at intermediate $q$. However, it still lacks a correct modeling of electron wavefunctions near the nucleus, whose contribution is dominant for large $q$. Having in mind that for energy deposition in the keV range $q$ can be up to 50 MeV for $m_\chi$ in the GeVs, it is exceptionally important to include relativistic effects in the $f_{ion}$ computation of this COSINE-100 data analysis of the DM-electron scattering model.

\section{Determination of constraints upper bounds due to Y2L and Yemilab overburden}\label{sec:appB}
Before reaching the detectors in underground experiments, DM particles may scatter in the atmosphere and surrounding rock overburden. This Earth shielding effect can attenuate the flux of DM at the detector, limiting the excluded parameter space at larger cross sections. As a result, sub-GeV DM searches commonly report upper bounds on the derived constraints for both DM–electron \cite{supercdmswimpelectron,damicmwimpelectron} and DM–nucleon \cite{cresstwimpnucleon,tesseractwimpelectron} studies. This effect is particularly relevant for DM–electron scattering searches, since for the cross sections accessible to direct detection experiments the impact of Earth shielding cannot be neglected.

The Verne 2.1 code \cite{vernegit,verneheavy,vernelight} has been used to compute the DM velocity distribution after the particles propagate through the laboratory overburden. A dark photon mediator has been assumed, as well as DM scattering with nucleons, for which the cross section is expected to be significantly larger than ${\overline\sigma}_e$ for the DM masses considered \cite{wimpnucleonandwimpelectron}. The laboratory depths are fixed to 700 m for Y2L and 1000 m for Yemilab. Also, the default Verne rock and atmospheric compositions have been adopted.

Several simplifying assumptions have been adopted in the evaluation of the Earth shielding effect. DM particles have been considered to reach the laboratories at an angle of 54° off the vertical. Although this angle varies during an year and a day, 54° corresponds to the annual average incidence angle. A flat rock overburden has been assumed, neglecting the detailed mountain profiles above Y2L and Yemilab. Finally, the default Verne rock composition has been taken to be uniform along the entire DM trajectory.

To obtain upper bounds on the constraints, the expected energy spectra in the COSINE-100 and COSINE-100U crystals after including the detector response have been generated using the DM velocity distributions calculated with Verne. The upper bound values of the DM–electron cross section are then defined by requiring the total number of events in the ROI to match that obtained for the cross sections that define the constraints in Figs. \ref{fig:limits} and \ref{fig:cosine100U}.

It is important to highlight that Verne 2.1 includes a module designed for calculating the Earth-shielding effect for light DM particles \cite{vernelight}, and a separate module appropriate for heavy DM particles \cite{verneheavy}. For intermediate DM masses, approximately in the 10 MeV $\sim$ 1 GeV range, which is considered in this work, care must be taken in interpreting the resulting upper bounds, since the approximations employed in both modules may become less accurate. Nevertheless, the upper bounds derived in this analysis lie at least one order of magnitude above the defined constraints and should be interpreted as naive estimates of the Earth shielding effect at Y2L and Yemilab.



%

\end{document}